\documentclass[12pt`]{article}

\newcommand  {\Rbar} {{\mbox{\rm$\mbox{I}\!\mbox{R}$}}}

\newsavebox{\zzzbar}
\sbox{\zzzbar}
  {\setlength{\unitlength}{0.9em}
  \begin{picture}(0.6,0.7)
  \thinlines
  \put(0,0){\line(1,0){0.6}}
  \put(0,0.75){\line(1,0){0.575}}
  \multiput(0,0)(0.0125,0.025){30}{\rule{0.3pt}{0.3pt}}
  \multiput(0.2,0)(0.0125,0.025){30}{\rule{0.3pt}{0.3pt}}
  \put(0,0.75){\line(0,-1){0.15}}
  \put(0.015,0.75){\line(0,-1){0.1}}
  \put(0.03,0.75){\line(0,-1){0.075}}
  \put(0.045,0.75){\line(0,-1){0.05}}
  \put(0.05,0.75){\line(0,-1){0.025}}
  \put(0.6,0){\line(0,1){0.15}}
  \put(0.585,0){\line(0,1){0.1}}
  \put(0.57,0){\line(0,1){0.075}}
  \put(0.555,0){\line(0,1){0.05}}
  \put(0.55,0){\line(0,1){0.025}}
  \end{picture}}
\newcommand{\Zbar}{\mathord{\!{\usebox{\zzzbar}}}}

\newcommand{\Z}{\Zbar}
\newcommand{\R}{\Rbar}

\begin{document}

\setlength{\textheight}{21cm}

\title{{\bf STATISTICAL MECHANICS OF ENTROPY PRODUCTION:\\
Gibbsian hypothesis and local fluctuations}}

\author{{\bf Christian
Maes}\thanks{Email: Christian.Maes@fys.kuleuven.ac.be }
\\ Instituut voor
Theoretische Fysica
\\ K.U.Leuven, B-3001 Leuven, Belgium.}

\maketitle

\begin{abstract} It is argued that a Gibbsian formula for the space-time distribution of
microscopic trajectories of a nonequilibrium system provides a unifying
framework for recent results on the fluctuations of the entropy
production.  The variable entropy production is naturally expressed as
the time-reversal symmetry breaking part of the space-time action
functional.  Its mean is always positive. This is both supported by a
Boltzmann type analysis by counting the change in phase space extension
corresponding to the macrostate as by various examples of nonequilibrium
models.  As the Gibbsian set-up allows for non-Markovian dynamics, we
also get a local fluctuation theorem for the entropy production in
globally Markovian models. In order to study the response of the system to
perturbations, we can apply the standard Gibbs formalism.
\end{abstract}
\vspace{3mm} \noindent

PACS number(s): 05.40.-a, 05.70.Ln, 65.40.Gr

\section{Introduction}
A rough and ready distinction between nonequilibrium phenomena can be made
by asking whether the system is in a transient state relaxing towards
equilibrium or whether the system is in a steady nonequilibrium state.
That can depend on the level of description and on the relevant length,
time and energy scales.  In what follows, we mostly have in mind the ideal
steady state scenario where a spatially extended system is driven away
from equilibrium via some external forcing or via contacts with unequal
reservoirs through which a current is maintained in the system. This is a
vast area of research with many beautiful models and results but it is
also  characterized by a lack of general guiding principles.  This in
contrast with equilibrium statistical mechanics where elementary questions
(such as to do with the positivity of certain response functions, their
relation with fluctuations, weak coupling expansions, etc.) have a
general answer mostly because of the power and the generality of the
underlying Gibbs formalism.  The next section proposes to extend the
domain of application of this Gibbs formalism to nonequilibrium
situations.  First we will state the main ideas in a somewhat formal way
and then we will discuss examples and some of the consequences for
typical models of nonequilibrium statistical mechanics.  This paper is a
continuation of \cite{M}; we have discussed other examples and general
consequences in \cite{M2,M3,M4,mmj,M5}. The main motivation and source of
inspiration have been on the one hand to try to understand the work of
Gallavotti-Cohen in \cite{gc1,gc2} on a general symmetry in the
fluctuations of the phase space contraction rate for certain strongly
chaotic reversible dissipative dynamical systems, in particular, to
interpret their so called chaoticity hypothesis, and on the other hand to
apply the theory of space-time Gibbs measures that was started more than
a decade ago for probabilistic cellular automata and that has since then
been developed way beyond, \cite{bk,daipra,gm,pca,M1,min2}.

\section{Main ideas}
\subsection{Gibbsian hypothesis}
By definition, in equilibrium, every region is in equilibrium with the
rest of the system.  If we imagine a subregion in our big equilibrium
system, then we will see there the same equilibrium state as outside for
the same microscopic interaction. Mathematically, this can be readily
seen from the form $ \exp[-\beta H]$ for equilibrium distributions (here
in thermal equilibrium at inverse temperature $\beta$ for the local and
additive microscopic interaction Hamiltonian $H$).  To be specific, think
of the Ising model in the square  $V$ and compute the expectation of a
function $f$ that only depends on the spins $\sigma_\Lambda$ inside the
smaller square $\Lambda$:
\begin{equation}\label{ha}
\langle f(\sigma_\Lambda) \rangle_V = \langle \frac
1{Z_{\Lambda}(\beta,\sigma_{\partial \Lambda})} \sum_{\sigma_\Lambda}
f(\sigma_\Lambda) e^{-\beta H_\Lambda^{\sigma_{\partial
\Lambda}}(\sigma_{\Lambda}) }\rangle_V
\end{equation}
with $\sigma_{\partial\Lambda}$ the spin configuration on the external
boundary of $\Lambda$. To appreciate is that the Hamiltonian
$H_\Lambda^{\sigma_{\partial \Lambda}}(\sigma_\Lambda) =
H_\Lambda(\sigma_\Lambda) +
W_{\partial\Lambda}(\sigma_\Lambda,\sigma_{\partial\Lambda})$
 is just changed
at the boundary of $\Lambda$: $H_\Lambda$ contains the bulk contribution
and $W_{\partial\Lambda}$ contains the interaction of the spins inside
$\Lambda$ with those outside; for any two boundary conditions
$\sigma_{\partial \Lambda}$ and $\sigma_{\partial \Lambda}',
|H_\Lambda^{\sigma_{\partial \Lambda}} - H_\Lambda^{\sigma_{\partial
\Lambda}'}| \leq c|\partial \Lambda|$. Moreover, upon specifying the
boundary condition $\sigma_{\partial \Lambda}$ the inside and the outside
of the region are decoupled. A more general formulation of this is known
as the Dobrushin-Lanford-Ruelle equation, see e.g. \cite{vE}, pages
891--956, for a mathematical introduction to the Gibbs formalism in
equilibrium statistical mechanics.\\ Now to nonequilibrium steady states.
There is no reason to expect that in general the stationary state, that is
the time-invariant distribution, can be described via a sufficiently
local Hamiltonian. Even if it is, there does not seem to exist a
systematic way of finding it. Typically, if we apply a time-independent
constraint (like fixing a rod in running water), we will, unlike in
equilibrium, greatly perturb the system. Think however of letting the rod
go with the water. It will do whatever the currents are telling it to do.
In order to specify these currents, or, to say how fast what amount of a
certain quantity is transported to where, we need space-time. Thus take a
space-time window $\Lambda\times [-\tau, \tau]$. If we specify at its
borders whatever went inside or outside at what pace, we can expect that
the inside is effectively decoupled from the outside, just as in
equilibrium. We conclude that the space-time distribution of
nonequilibrium steady states may well be sharing the Gibbsian features of
equilibrium states.\\ It is interesting to compare the situation with
what happens in the low temperature phases of the two-dimensional Ising
model. We can think of it as a space-time distribution corresponding to a
`dynamics' formally specified by the transfer matrix. But the restriction
of say the low temperature plus phase to a one-dimensional layer is not a
Gibbs distribution for any uniformly and absolutely summable interaction
potential, see \cite{Sch}. That would correspond to the time-invariant or
stationary state. There are ways here to restore its Gibbsian
characterization by slightly extending the definition of Gibbs measure
(as was for example done in \cite{MRV2}) but it remains unclear what
aspects of the Gibbsian formalism remain valid and, more to the point,
there does not seem an easy way to generalize this in any systematic way.
One could object that this Ising example has a major defect as its
`dynamics' (imposed by the transfer matrix) is highly nonlocal. There are
other examples for a strictly local dynamics, such as in \cite{LSc} for
the voter model, where the invariant measures are not Gibbsian. Yet, the
point is that while it is mostly easily seen that the space-time
distribution is Gibbsian for some local space-time action functional that
can be explicitly constructed from the dynamics, it is in general very
hard to get the same thing for the projection on a fixed time layer, that
is, for the corresponding invariant distribution. Another mathematical
analogy is to think of the difference between the Hamiltonian and the
Langrangian set-up in mechanics.

We now add some formulas. We consider a system in a volume $V$ which is
observed over a time-interval $[-\tau, \tau]$. At each time, we have the
same microscopic phase space $K$ and a trajectory is denoted by $\omega =
(\omega_t, t\in [-\tau,\tau]), \omega_t\in K$; $\omega$ is the history of
the microscopic degrees of freedom (for example containing
 the positions and momenta of all the particles in a box over a certain
time-interval). We prepare the system at time $ -\tau$ in some macrostate
corresponding to a probability distribution $\rho$ on $K$.
 We write $P=P_\rho^\tau$ for the probability
distribution of the trajectories $\omega$ given that it started as
sampled from $\rho$. How to get this distribution $P$ is another question
on which we will soon come back below.  Under the Gibbsian hypothesis we
assume that $P_\rho^\tau$ has a density with respect to a corresponding
equilibrium distribution $P_o=P_{o,\rho^o}^\tau$ given as
\begin{equation}\label{af}
dP(\omega) = e^{-A(\omega)} \, dP_o(\omega)
\end{equation}
 where $A(\omega)$ is the space-time action-functional;
it will contain space-time integrals over the microscopic trajectories.
The nonequilibrium distribution $P$ is obtained from the equilibrium
distribution $P_o$ by adding some external forces or couplings to
reservoirs with different thermodynamic parameters.  There are various
ways then to obtain (\ref{af}). Given a particular dynamics describing an
open system it is in general easy to construct $A$ explicitly (modulo
space-time boundary terms) and this is what we will illustrate in the
next section.  We get the form
\begin{equation}\label{ham}
A(\omega) = \ln \rho_0(\omega_{-\tau}) - \ln \rho(\omega_{-\tau}) + {\cal
H}(\omega)
\end{equation}
where $\rho, \rho^o$ are the initial data (at time $-\tau$) for the
nonequilibrium, respectively, the equilibrium process and the form of
${\cal H}$ depends only on the dynamics. Most important is that ${\cal
H}$ is local, and approximately additive for a spatially extended local
dynamics.  In this way, ${\cal H}$ is like a Hamiltonian in equilibrium
statistical mechanics but here for space-time trajectories. In case we
consider a steady state process, then $\rho$, respectively, $\rho^o$
should be taken as stationary (i.e., time-invariant) distributions under
the nonequilibrium and the equilibrium dynamics.  It may seem problematic
to assume in general that stationary states have a density with respect to
some {\it a priori} flat measure, i.e., that they are absolutely
continuous with respect to some analogue of the Lebesgue measure on $K$.
In fact, all that is really required here is a local absolute continuity
with respect to the Lebesgue measure; since we have in mind macroscopic
systems where the macrovariables correspond to sums of local functions,
we have every reason to expect that the distribution on $K$ inferred from
their values will indeed have a density.\\
On the other hand, from the formula (\ref{af}), it is suggestive to
forget altogether about the specific dynamics and to concentrate on $A$
instead.  Each $A$ gives rise to some type of dynamics but it need not be
Markovian.  The specific form of $A$ then defines the nonequilibrium
model.

The connection between Gibbs measures and nonequilibrium dynamics is not
new; the theory has been worked out into considerable detail for various
situations, see \cite{pca,min2,M,M1,daipra,bk,gm} for examples.

\subsection{Time-reversal}
We are used to forget about time when dealing with equilibrium statistical
mechanics.  One reason is that, for classical systems, the momenta
(entering only in the kinetic energy part of the Hamiltonian) can be
integrated out at once from the partition function. Time enters more
explicitly in equilibrium dissipative dynamics such as via Langevin
equations  or Glauber dynamics. Yet, under the condition of detailed
balance, the stationary dynamics is microscopically reversible and the
past cannot be distinguished from the future. One essential feature of
nonequilibrium is that time-reversal invariance is
broken.\\
The time-reversal operation $\Theta$ acts on path space and it consists
of two parts. The first part $\pi$ is purely kinematical and depends on
the nature of the dynamical variables; the second part is just the
reflection of the time-axis $t\rightarrow -t$. So if $\omega= (\omega_t,
t\in [-\tau,\tau])$ is a trajectory, then its time-reversal is $\Theta
\omega$ with $(\Theta \omega)_t = \pi\omega_{-t}$ where $\pi$ is an
involution on phase space (for example reversing the sign of the momenta).
Since the violation of microsopic reversibility is so essential to the
presence of currents and nonequilibrium conditions, it must be clear that
important information must be written in the symmetry breaking part
$\Delta {\cal H}(\omega) \equiv {\cal H}(\Theta \omega) - {\cal
H}(\omega)$ of the space-time Hamiltonian (\ref{ham}).  Or, in terms of
the Gibbsian distribution $P=P_\rho^\tau$ of (\ref{af}),
\begin{equation}\label{r}
dP_\rho^\tau(\omega) = e^{R^\tau_\rho(\omega)} \,
dP_{\rho_\tau\pi}^\tau\Theta(\omega)
\end{equation}
We have the notation $P_{\rho_\tau\pi}^\tau$ for the distribution of the
trajectories started at time $-\tau$ in the state $\rho_\tau\pi$ where
$\rho_\tau$ corresponds to the macrostate in which we ended up at time
$\tau$ when started at time $-\tau$ in $\rho$.  More explicitly, from
(\ref{ham}),
\begin{equation}\label{echt}
R_\rho^\tau(\omega) = -\ln
\frac{\rho_\tau(\omega_\tau)}{\rho^o(\omega_\tau)} + \ln
\frac{\rho(\omega_{-\tau})}{\rho^o(\omega_{-\tau})} + \Delta {\cal
H}(\omega)
\end{equation}
It is physically more convenient to decompose $\Delta {\cal H}$ a bit
further and to write
\begin{equation}\label{pmc}
R_\rho^\tau(\omega) = -\ln \rho_\tau(\omega_\tau) + \ln
\rho(\omega_{-\tau}) + \Delta S_e
\end{equation}
where $\Delta S_e$, as we will illustrate, corresponds to the change of
entropy in the external world (baths).  We will see it in the explicit
formula (\ref{rheat}) in example 3.2. As for ${\cal H}$, also $\Delta
{\cal H}$ will be a space-time integral of local interaction
terms.\\
This object  has some remarkable properties.  In fact we will argue in
the next section
 via examples and in Section 4 via some
theoretical considerations that $R_\rho^\tau$ is the total variable
(space-time-integrated) entropy production but before, there are some
simple mathematical facts that are worth observing. First of all, its
mean equals
\begin{equation}\label{rt}
R^\tau(\rho) \equiv \int dP_\rho^\tau \ln
\frac{dP_\rho^\tau}{dP_{\rho_\tau\pi}^\tau \Theta} \geq 0
\end{equation}
That is a relative entropy: the path space expectation of the logarithm of
the density of that same path space measure with respect to the
time-reversed process started at $\rho_\tau$. We can also look at
(\ref{rt}) as a functional on the fixed time distributions $\rho$. By
making $\tau$ very small, we then get the entropy production in the
transient state $\rho$. It is thus natural to define the mean entropy
production rate in $\rho$ as
\begin{equation}\label{take1}
\dot{S}(\rho) \equiv \frac 1{2} \frac{dR^\tau(\rho)}{d\tau}\,(\tau=0)
\end{equation}
We will take up this discussion in Section 4, at formula (\ref{takeup}).
Let us now continue for a while with a steady state process (where
$\rho=\rho_\tau$ is a stationary measure).  In this case we write
$R^\tau_\rho(\omega)=R(\omega)$ and $\dot{S}(\rho)= \langle R \rangle
/(2\tau)$ is obtained via
\begin{equation}\label{pos}
\langle R \rangle= \int dP(\omega) R(\omega) = S(P|P\Theta) \geq 0
\end{equation}
where the right hand side is the relative entropy between the forward and
the backward distribution: formally (and in most cases to be understood
via a sequence of space-windows),
\[
s(P|P\Theta) \equiv \int dP(\omega)\ln \frac{dP}{dP\Theta}(\omega)
\]
It is this quantity that turns out to be the steady state mean entropy
production: it
 is always non-negative and is zero if and only
if the dynamics is microscopically reversible (detailed balance). Of
course for spatially extended systems $R$ is an extensive quantity and
when divided by the spatial volume, it will take on a definite limiting
value as a consequence of the law of large numbers.
So its positivity is typical and not only an averaged property.\\
 For the
fluctuations  we have, almost immediately since $R(\Theta\omega)
=-R(\omega)$,
\begin{equation}\label{flu}
\int dP(\omega) e^{-z R(\omega)} = \int dP(\omega) e^{-(1-z) R(\omega)}
\end{equation}
for all complex numbers $z$.  This is both for its derivation and for its
possible applications very much like a Ward identity (with respect to a
discrete symmetry), see e.g. \cite{Sim}. A similar symmetry for the
fluctuations of the time-averaged phase space contraction in the so
called SRB-measure for a class of strongly chaotic dynamical systems was
first observed in \cite{ecm} and derived in \cite{gc1,gc2,Gen,Ru4}.
  Upon
differentiating (\ref{flu}) with respect to $z$ and with respect to the
parameters hidden in the distribution $P$, we obtain exact relations
between space-time correlations.  Such derivations have already been
explored in \cite{G,LeS,M}.  This would not be very useful if it were not
that the same symmetry remains valid also for {\it local} fluctuations.
The search for a local fluctuation theorem already started in
\cite{cl,G3,M,GP} and a systematic theoretical answer was developed in
\cite{mmj}.  It is the Gibbsian set-up that saves us: if we take the
restriction $P_\Lambda$ of $P$ to a space-time window $\Lambda\times
[-\tau,\tau]$ and if we apply the time-reversal $\Theta_\Lambda$ only
there, then, by definition, for all functions $f$ of the trajectory in
$\Lambda\times [-\tau,\tau]$
\begin{equation}\label{locr}
\int dP(\omega) f(\Theta \omega) = \int dP(\omega) f(\omega)
e^{-R_\Lambda(\omega)}
\end{equation}
where \[ R_\Lambda(\omega) \equiv \ln \frac{dP_\Lambda}{dP_\Lambda
\Theta_\Lambda}(\omega) \] and $P_\Lambda\Theta_\Lambda =
(P\Theta)_\Lambda$. Therefore, by substituting $f(\omega) = \exp[z
R_\Lambda(\omega)]$ in (\ref{locr}), we get a local version of
(\ref{flu}). The important point is now that, due to the locally additive
character of $R$, its restriction to $\Lambda$ is exactly equal, modulo
space-time boundary terms, to $R_\Lambda$ so that we really get control
over the local fluctuations of the entropy production. The only
assumption here is that whenever the restriction of a trajectory $\omega$
to $\Lambda\times [-\tau, \tau]$ has positive probability to occur, then
the restriction of its time-reversal $\omega\Theta$ to $\Lambda\times
[-\tau, \tau]$ is also possible (has positive probability). This is a
condition of dynamic reversibility which should not be confused with
microscopic reversibility under which the two probabilities would be
exactly
equal!\\
We can continue a bit more with (\ref{locr}). Suppose we condition the
left hand side on a particular value $r$ of the entropy production \[
\langle f\Theta | R_\Lambda =r \rangle = \langle f | R_\Lambda = -r
\rangle \] so that, for fixed entropy production in $\Lambda$, the
time-reversal operation can be exchanged with conditioning on the
opposite entropy production. The local arrow of time is thus decided by
the sign of the local entropy production.  A similar point was made for
the phase space contraction
in strongly chaotic dynamical systems in \cite{G4}.\\
Finally, another consequence of (\ref{locr}) is that by Jensen's
inequality, when $f$ is positive,
\begin{equation}\label{jens} \langle R_\Lambda
\rangle_f \geq \ln \frac{\langle f\rangle}{\langle f\Theta\rangle}
\end{equation} where $\langle R_\Lambda \rangle_f \equiv \langle f
R_\Lambda \rangle / \langle f\rangle$ is the expected local entropy
production when $P$ is perturbed by the insertion of an extra density
$f$. One choice takes $f = R_\Lambda^n$ with $n$ even, for which
(\ref{jens}) implies that $\langle R^{n+1}_\Lambda \rangle \geq 0$ for
all $n$.\\ Similar relations can be derived for the transient entropy
production. We restrict us here to the analogue of (\ref{flu}) for $z=1$.
We have with no extra effort, upon substituting (\ref{pmc}),
\begin{equation}\label{flut}
\int dP_\rho^\tau(\omega) e^{-[ -\ln \rho_\tau(\omega_\tau) + \ln
\rho(\omega_{-\tau}) + \Delta S_e(\omega)]} = 1
\end{equation}
which expresses just the normalization of the distribution
$P_{\rho_\tau\pi}^\tau \Theta$. An interesting possibility is that
$\rho(\eta)=\exp -\beta H_i(\eta)/Z_i$ and $\rho_\tau(\eta)=\exp -\beta
H_f(\eta)/Z_f$ are initial and final equilibrium states at the same
inverse temperature $\beta$ but for a different Hamiltonian.  As a
physical mechanism we can consider the system coupled to a heat bath at
constant temperature $T=1/\beta$ where some parameters (e.g. interaction
coefficients) in the interaction of the components of the system are
changed. This means that the Hamiltonian $H(t) \equiv H(\lambda(t),\eta)$
is time-dependent and $H_f(\eta) \equiv H(\lambda(\tau),\eta), H_i(\eta)
\equiv H(\lambda(-\tau),\eta)$.  To change the parameter $\lambda$ from
$\lambda(-\tau)$ to $\lambda(\tau)$ some heat must flow from the bath
into the system so that the change of entropy of the bath equals $\Delta
S_e = -\beta[H_f(\eta_{\tau}) - H_i(\eta_{-\tau}) - W_\tau]$ where
 $W_\tau$ is the irreversible work
done over the time $[-\tau,\tau]$. The identity (\ref{flut}) then becomes
\[
\frac{Z_f}{Z_i} = \int dP_\rho^\tau(\omega) e^{-\beta W_\tau(\omega)}
\]
and the left hand side is the exponential of a change in equilibrium free
energy. A similar relation was discussed in \cite{c,c1,hs,j,M4} where it
is also connected with a
fluctuation identity for the transient nonequilibrium regime, \cite{ecm2,gc3}.\\
The next section illustrates some of the above for some typical models of
nonequilibrium statistical mechanics.

\section{Examples}
We concentrate here on some aspects of two standard models of
nonequilibrium physics.  More examples of parts of the theory above can
be found in \cite{M4}.

\subsection{Reaction-Diffusion process}
In a microscopic version of a reaction-diffusion system, particles can
disappear or be created on the sites of a regular lattice and they can
hop to
nearest neighbor vacancies; see \cite{LIGG} for a technical introduction.\\
We consider variables $\eta(i) = 0,1$ on the sites $i$ of the square
lattice $\Z^2$. We interpret it as meaning that site is empty or occupied
by a particle. The phase space is $K=\{0,1\}^V$ where $V$ is a large
square of side length $N$ centered around the origin. The dynamics
consists of a reaction part: $\eta\rightarrow \eta^i$ where $\eta^i$ is
identical to $\eta$ except that the occupation at the site $i$ is flipped
and a diffusion part: $\eta\rightarrow \eta^{ij}$ where $\eta^{ij}$ is
the new configuration obtained by exchanging the occupations at sites $i$
and $j$; we take periodic boundary conditions on $V$. We get a driven
lattice gas by adding an external field $E>0$ giving a bias for particle
hopping in a certain direction. \\ In formula: first, the probability per
unit time to flip from a configuration $\eta$ to the new $\eta^i$ is
\[
c(i, \eta) = \gamma_+ (1-\eta(i)) + \gamma_- \eta(i)
\]
where $\gamma_+$ is the rate for the transition $0\rightarrow 1$ and
$\gamma_-$ is the rate for $1\rightarrow 0$. Secondly, the hopping rates
over a nearest neighbor pair $\langle ij\rangle$ in the horizontal
direction, $i=(i_1,i_2), j=(i_1+1,i_2)$:
\[
c(i, j, \eta) = e^{E/2} \eta(i)(1-\eta(j)) + e^{-E/2} \eta(j)(1-\eta(i))
\]
The hopping rate in the vertical direction is symmetric (put $E=0$ in the
above if $j=(i_1,i_2\pm 1)$). Taking $E$ large, typically, many more
particles will be jumping to the right than to the left. In the absence of
reaction rates, that is for $\gamma_{\pm}=0$, we recover the so called
asymmetric exclusion process and particle number is strictly conserved.
The stationary state $\rho$ is a product state with uniform density equal
to $\gamma_+ / (\gamma_- + \gamma_+)$ corresponding to a chemical
potential $\ln \gamma_+/\gamma_-$ of the particle
reservoir.\\
As reference equilibrium process $P_o$, we take the same model with $E=0$
and we can take $\rho^o=\rho$. Consider now a trajectory
$\omega=(\omega_t, t\in [-\tau, \tau])$. It is then easy to find
(\ref{af}) with
\begin{eqnarray}
A(\omega) &=& -\frac{E}{2} \sum_{i\in V} \sum_{-\tau\leq t\leq \tau}
\omega_t(i)(1-\omega_t(j)) - \omega_t(j)(1-\omega_t(i))
\\\nonumber +&& \int_{-\tau}^\tau dt \sum_{i\in V}
(e^{E/2}-1)\omega_t(i)(1-\omega_t(j)) +
(e^{-E/2}-1)\omega_t(j)(1-\omega_t(i))
\end{eqnarray}
where $j=(i_1+1,i_2)$, the right neighbor of $i=(i_1,i_2)$ and the first
sum is over all jump times over the bond $\langle ij\rangle$ in the
trajectory $\omega$. Applying the time-reversal
$(\Theta\omega)_t=\omega_{-t}$ we compute that here
\begin{equation}\label{work}
R(\omega)=\Delta {\cal H}(\omega) =  E \sum_{i\in V} J_\tau^{i}(\omega)
\end{equation}
where $J_\tau^{i}$ is the time-integrated microscopic current over a
fixed nearest neighbor pair $\langle ij\rangle$, with $j= (i_1+1,i_2)$ to
the right of $i$.  Explicitly,
\[
J_\tau^{i}(\omega)) \equiv \sum_{-\tau\leq t\leq \tau}
[\omega_{t}(i)(1-\omega_t(j)) - (1-\omega_t(i))\omega_t(j)] \] where
again the sum is over all jump times $t$ for the bond $\langle
ij\rangle$.  In other words $\Delta {\cal H}$ is the variable work done
on our system by the external field over the time-interval $[-\tau,
\tau]$. Its average in the stationary state equals (up to a temperature
factor) the heat dissipated in the environment. In fact we can compute
this:
\[
\langle \Delta S_e \rangle = \langle \Delta {\cal H}\rangle = 4\tau|V| \,
E \sinh(E/2) \, \frac{\gamma_+ \gamma_-}{(\gamma_+ + \gamma_-)^2}>0
\]
Let us now fix a rectangle $\Lambda=\{(i_1,i_2)\in V:
i_1=1,\ldots,\ell,i_2=1,\ldots,n\}$ of height $n$ and width $\ell$. To
study the fluctuations of the work restricted to $\Lambda$, we follow the
set-up of Section 2, equation (\ref{locr}), and we introduce the local
time-reversal operator $\Theta_\Lambda$ for which
$(\Theta_\Lambda\omega)_t(i)=\omega_{-t}(i)$ if $i\in \Lambda$; the other
occupation variables are left unchanged. To compute the $R_\Lambda$ of
(\ref{locr}) we note that the restriction $P_{E,\Lambda}$ of the steady
state $P=P_E$ to the volume $\Lambda$ satisfies
\[
P_{E,\Lambda}\Theta_\Lambda = (P_{E}\Theta)_{\Lambda}=P_{-E,\Lambda}
\]
so that (\ref{locr}) is verified with
\[
R_\Lambda = \ln \frac{dP_{E,\Lambda}}{dP_{-E,\Lambda}}
\]
Here is the bulk of the formula
\begin{equation}\label{lft}
R_\Lambda =  E \sum_{i\in \Lambda: (i_1+1,i_2) \in \Lambda} J_\tau^{i}
\pm c(n+\ell)\tau
\end{equation}
We have not written out the correction term; we refer to \cite{mmj} for
details and proofs;  the important thing is that this correction is of
the order of the boundary of $\Lambda$ times $\tau$ and the factor $c$
depends on the field $E$ and on the chemical potential but it is
independent of the big volume $V$. The first term in $R_\Lambda$ is of
the order of the volume and it is the variable work done on the system in
$\Lambda$ over the time-interval $[-\tau, \tau]$.   The main contribution
to $R_\Lambda$ is thus exactly the restriction of (\ref{work}) to
$\Lambda$, as announced in Section 2 (following the identity
(\ref{locr})).\\
For the proof of the resulting local fluctuation theorem we refer to
\cite{mmj}.  There is one aspect, the analogue of (\ref{ha}), that is
essential. Suppose we want to compute the expectation at time $t$ of a
function $f$ that only depends on the particle configuration in $\Lambda$
given the history of that configuration. That is \[ \langle
f(\omega_t(i), i\in \Lambda) | \omega_{s}(j)=\eta_s(j), j\in \Lambda,
s\leq t-\delta\rangle
 \]
  Clearly, the evolution restricted to $\Lambda$ is
no longer Markovian, but what are the rates for the elementary
transitions, that is taking $\delta$ very small in the above expectation?
The answer is that all the rates remain unchanged except at the boundary.
So the restricted process has the same transition rates in the bulk of
$\Lambda$ just as we saw before in ({\ref{ha}) that in equilibrium the
Hamiltonian is unchanged for the restriction to $\Lambda$.

\subsection{Nonequilibrium crystal}
The previous example was a bulk driven lattice gas.  We now take a
surface (thermally) driven Hamiltonian system. For large integer $N$ we
consider a linear crystal $L_N \equiv \{-N,\ldots,0,\ldots,N\}$ where
each site $i\in L_N$ carries a particle with momentum $p_i$ and position
$q_i$ (real variables).  The Hamiltonian is
\begin{equation}\label{hamt}
H_N(p,q) \equiv \frac 1{2} \sum_{i=-N}^N  p_i^2 + U_N(q)
\end{equation}
with a nearest neighbor potential
\[
U_N(q) = \sum_{i=-N}^N V_i(q_i) + \sum_{i=-N}^{N-1} \lambda_i \Phi(q_i -
q_{i+1}).
\]
Here we will not care about exposing conditions on the potential which
are needed for what follows; the
harmonic crystal would correspond to a quadratic potential.\\
The proposed dynamics is stochastic and generated by Hamilton's equations
of motion in the bulk and by Langevin equations at the boundaries:
\[
dq_i = p_i dt, i\in L_N
\]
\[
dp_i = -\frac{\partial U_N}{\partial q_i}(q) dt, i=-N+1,\ldots,N-1
\]
\[
dp_{-N} =-\frac{\partial U_N}{\partial q_{-N}}(q) dt -\gamma p_{-N} dt +
\sqrt{\frac{2\gamma}{\beta_{\ell}}} dW_{\ell}
\]
\[
dp_{N} =-\frac{\partial U_N(q)}{\partial q_{N}}(q) dt -\gamma p_{N} dt +
\sqrt{\frac{2\gamma}{\beta_{r}}} dW_{r}
\]
where $\gamma >0$ and $\beta_\ell$, respectively $\beta_r$, are the
inverse temperatures at the left and the right end of the chain. $W_\ell$
and
$W_r$ are independent standard Wiener processes (It\^o sense).\\
$\mathcal{L}_N$ is the generator of the dynamics:
\[
\mathcal{L}_N \equiv p\cdot\nabla_q - \nabla_q U_N \cdot \nabla_p -\gamma
p_{-N} \frac{\partial}{\partial p_{-N}} -\gamma p_{N}
\frac{\partial}{\partial p_{N}} + \frac{\gamma}{\beta_\ell}
\frac{\partial^2}{\partial p_{-N}^2} + \frac{\gamma}{\beta_r}
\frac{\partial^2}{\partial p_{N}^2}
\]
Note that if $\beta_\ell=\beta_r=\beta$, then the Gibbs measure on
$\R^{4N+2}$ with density
\[
\rho^o(p,q) \equiv \frac 1{Z_N} e^{-\beta H_N(p,q)}
\]
is reversible in the sense that when $\beta_\ell=\beta_r=\beta$, then
$\mathcal{L}_N^\ast = \pi \mathcal{L}_N \pi$ on
$L^2(\rho^o)$ where $\pi f (p,q) \equiv f(-p,q)$.\\
We are interested in  the case $\beta_\ell < \beta_r$.  We assume that
the interaction $U_N$ is sufficiently well behaved to allow for a unique
stationary measure with a smooth density with respect to $dpdq$. To
actually prove this, requires some specific assumptions on the potential;
here we take this for granted and we refer to \cite{eck1,eck2} for
mathematical results.  We denote this unique stationary measure by
$\rho$. The corresponding (stationary) path space measure $P$ in the
time-interval $[-\tau,\tau]$ is the law of the stationary Markov
diffusion process $\omega\equiv (\omega(t)\equiv (p(t),q(t)), t\in
[-\tau,\tau])$ described by the dynamics above with invariant measure
$\rho$. In this nonequilibrium steady state heat will flow from the left
end (hot) to the right (cold) and entropy will be produced.   This
nonequilibrium model has frequently appeared, see \cite{N,RLL} for the
harmonic
crystal, and see \cite{eck1,eck2} for anharmonic examples.\\
In order to describe the process $P$ via a space-time action functional
$A$, as in (\ref{af})-(\ref{ham}), we need a reference process.  In fact
there are many choices. Consider the dynamics
\[
dq_i = p_i dt, i\in L_N
\]
\[
dp_i = -\frac{\partial U_N}{\partial q_i}(q) dt, i=-N+1,\ldots,N-1
\]
\[
dp_{-N} =-\frac{\partial U_N}{\partial q_{-N}}(q) dt -\gamma \kappa_{-N}
p_{-N} dt + \sqrt{\frac{2\gamma}{\beta_{\ell}}} dW_{\ell}
\]
\[
dp_{N} =-\frac{\partial U_N}{\partial q_{N}}(q) dt -\gamma \kappa_N p_{N}
dt + \sqrt{\frac{2\gamma}{\beta_{r}}} dW_{r}
\]
where the only difference with the original dynamics sits in the drift
term by the insertion of factors $\kappa_{-N}, \kappa_{N} > 0$.  If we
choose $\beta_{\ell}\kappa_{-N}=\beta_r \kappa_{N}=\beta$, then this
reference process $P_o$ is reversible with respect to the same stationary
measure $\rho^o$ as before with $\beta_\ell=\beta=\beta_r$. These laws
are mutually absolutely continuous and, applying the Girsanov formula
(taking it from \cite{LS}),
 we find for (\ref{af})-(\ref{ham}):
\begin{equation}\label{girs}
dP(\omega) =
\frac{\rho(\omega_{-\tau})}{\rho^o(\omega_{-\tau})}\,e^{-{\cal
H}(\omega)}\,dP_o(\omega)
\end{equation}
where
\begin{eqnarray}\label{H}
-{\cal H} (\omega) = \frac 1{2} \int_{-\tau}^\tau( (\beta-\beta_\ell)
 p_{-N}(t) dp_{-N}(t) &+& (\beta-\beta_r)
p_{N}(t) dp_{N}(t))\\\nonumber + \frac {\beta -\beta_\ell }{2}
\int_{-\tau}^\tau \frac{\partial U_N}{\partial q_{-N}}(q(t)) p_{-N}(t) dt
&+& \frac {\beta- \beta_r}{2}  \int_{-\tau}^\tau \frac{\partial
U_N}{\partial q_{N}}(q(t)) p_{N}(t) dt\\\nonumber +
\frac{\gamma}{4}\int_{-\tau}^\tau dt(
 (\beta \kappa_{-N} - \beta_\ell  ) p_{-N}(t)^2  &+&
(\beta \kappa_N -\beta_r ) p_{N}(t)^2)
\end{eqnarray}
Write $\Theta \omega$ for the time-reversal of the trajectory $\omega$,
that is: if $((p(t),q(t)), t\in [-\tau,\tau])$, then $\Theta \omega =
((-p(-t),q(-t)), t\in [-\tau,\tau])$.  In order to obtain
(\ref{r})-(\ref{pmc}), we must compute $ \Delta{\cal H} \equiv {\cal
H}\Theta - {\cal H}$, the relative action under time-reversal of $P$. The
result is
\begin{eqnarray}\nonumber
\Delta{\cal H}(\omega) = (\beta-\beta_\ell) && \int_{-\tau}^\tau
(p_{-N}(t) \circ dp_{-N}(t) + \frac{\partial U_N}{\partial q_{-N}}(q(t))
p_{-N}(t)
dt)\\
+(\beta-\beta_r) \int_{-\tau}^\tau (p_{N}(t) \circ dp_{N}(t) +&&
\frac{\partial U_N}{\partial q_{N}}(q(t)) p_{-N}(t) dt)
\end{eqnarray}
where the stochastic integral (indicated by the $\circ$) is now in the
Stratonovich-sense. Finally, we have (\ref{r})-(\ref{pmc}) with
\begin{eqnarray}\label{rheat}
&& R(\omega) = -\ln \rho(\omega_{\tau}) + \ln \rho(\omega_{-\tau})
-\beta_{\ell}[\frac 1{2} p_{-N}^2(\tau) - \frac 1{2}
p_{-N}^2(-\tau)\\\nonumber &&+ \int_{-\tau}^\tau \frac{\partial
U_N}{\partial q_{-N}}(q(t)) p_{-N}(t)dt] -\beta_r[\frac 1{2}
p_{N}^2(\tau) - \frac 1{2} p_N^2(-\tau) + \int_{-\tau}^\tau
\frac{\partial U_N}{\partial q_{N}}(q(t)) p_{N}(t)dt]
\end{eqnarray}
The first two terms give the change $\Delta S$ (between time $\tau$ and
time $-\tau$) of the entropy of the system.  The other terms give $\Delta
S_e$ of (\ref{pmc}).  Corresponding to the right end of the chain, we have
\begin{eqnarray}
&& \frac 1{2} p_{N}^2(\tau) - \frac 1{2} p_N^2(-\tau) + \int_{-\tau}^\tau
\frac{\partial U_N}{\partial q_{N}}(q(t)) p_{N}(t)dt = \\\nonumber &&
h_N(\omega_\tau) - h_N(\omega_{-\tau}) - \int_{-\tau}^\tau J^{+}(t) dt
\end{eqnarray}
where
\[
h_{N}(p,q) \equiv
 p_{N}^2/2 + V_{N}(q_{N})
 \]
 and $J^{+}$ is the flow of energy per unit time {\it into} the right
 reservoir:
 \[
 J^{+}(t) \equiv \lambda_{N-1} p_{N} \Phi'(q_{N-1}(t)-q_N(t))
 \]
and similarly for the current {\it into} the left reservoir:
\[
J^{-}(t) \equiv - \lambda_{-N} p_{-N} \Phi'(q_{-N}(t)-q_{-N+1}(t))
 \]
 and
 \[
h_{-N}(p,q) \equiv
 p_{-N}^2/2 + V_{-N}(q_{-N})
 \]
As required, for all $k\geq -N+1$,
\begin{eqnarray}\label{sst}
\frac{d}{dt} && \sum_{k}^{N-1}[\frac 1{2}p_i^2(t) + V_i(q_i(t))
+  \lambda_i \Phi(q_i(t)-q_{i+1}(t))]= J_k(t) - J^+(t)\nonumber\\
\mbox{and } && J_{-N+1}(t)= - J^{-}(t) - \lambda_{-N} \frac{d}{dt}
\Phi(q_{-N}(t)-q_{-N+1}(t))
\end{eqnarray}
with
\begin{equation}\label{cur}
J_k(t) \equiv \lambda_{k-1}\, p_k(t)\, \Phi'(q_{k-1}(t) - q_k(t))
\end{equation}
the current over the bond $k-1 \rightarrow k$, and $J^{+}(t)= J_{N}$.
Hence, the second contribution to the entropy production comes from the
energy transfers to the left and the right reservoirs due to the very
presence of the reservoirs (the energy of the system is not conserved):
it equals the sum of $-\beta_\ell \Delta h_{-N}\equiv -\beta_\ell
[h_{-N}(\omega_\tau)- h_{-N}(\omega_{-\tau}]$ and $-\beta_r \Delta
h_{N}\equiv -\beta_r [h_{N}(\omega_\tau)- h_{N}(\omega_{-\tau}]$ and this
will be part of $\Delta S_e$ even when
 the crystal is decoupled ($\Phi\equiv 0$).
  Finally, the
remaining third contribution consists of the heat current to the left and
to the right multiplied by the respective inverse temperatures. We can
then summarize (\ref{rheat}) as
\[
R(\omega) = \Delta S - \beta_\ell \Delta h_{-N} -\beta_r \Delta h_{N}
 + \beta_\ell \int_{-\tau}^\tau J^{-}(t) dt  +
\beta_r \int_{-\tau}^\tau J^{+}(t) dt
\]
From (\ref{sst}), in the steady state,  $-\langle J^{-}\rangle = \langle
J^{+}\rangle = \langle J_k\rangle $ and upon taking the steady state
average of (\ref{rheat}), as in (\ref{pos}), only this third contribution
\begin{equation}\label{poso}
2\tau (\beta_r - \beta_\ell) \langle J_0\rangle \geq 0
\end{equation}
survives and we see explicitly that the heat current goes from larger (at
the left) to smaller (at the right) temperatures.  While this result is
elementary from a thermodynamic perspective, the derivation above from a
statistical mechanical model, is, to our knowledge and taste, the
simplest and most natural one (compare e.g. with \cite{eck1,eck2,l1}).
Dividing the left hand side of (\ref{poso}) by the total time $2\tau$
gives the standard expression for the
mean entropy production rate $\dot{S}$.\\
This example will be continued in Section 5.3.

\section{Entropy production}

 In
 the previous scheme of Section 2.2, entropy production was identified with the time-reversal
 symmetry breaking part in the space-time action functional governing the
 nonequilibrium space-time distribution, see (\ref{r})-(\ref{pmc}).
 Of course, this is the {\it total} entropy production, that is the
 change of the entropy of system plus reservoirs.
   Looking at
 (\ref{pmc}) the first two terms give the change in entropy of the
 system and the rest corresponds to exchanges with the reservoirs or to external work.
 It is very well possible that in some transient regime, the entropy
 of the system {\it decreases} but the total entropy change is always
 positive, see ({\ref{rt}), (\ref{pos}).  In the steady state, entropy is
 constantly produced in the system but is carried away to the
 reservoirs. The mean entropy production $\langle R \rangle$ is always
time-reversal invariant (same entropy production in original as in
time-reversed steady state) and thermodynamic equilibrium is
characterized by zero mean entropy production (the absolute minimum).

 The easiest way to justify this mechanism
 is to look at classes of examples as in the above and just observe
 that it works.  For more examples, see \cite{M4}.  But there are also more
 general answers.\\
 The first one looks a bit easy but it is essential.
 Consider a closed system of particles subject to a Hamiltonian
 dynamics.  The entropy of a macrostate $M$
 is given via Boltzmann's formula $S(M)=\ln W(M)$ where $W(M)$ `counts' the number of
 microstates giving rise to the macrostate $M$; we refer to the recent
 \cite{Sh} for details. Suppose now that we condition on starting in this macrostate;
 what is the probability of a given microscopic trajectory?
 Since the trajectory is completely decided by the initial microstate, this probability is
 given by $1/W(M)$.  The trajectory ends in some microstate giving
 rise to a new macrostate $M'$.  The time-reversed trajectory must thus be conditioned
 on starting
 somewhere in the phase space corresponding to $M'$ and it has
 therefore a probability $1/W(M')$.  Hence, the ratio of these probabilities
 is  $W(M')/W(M) = \exp[S(M')-S(M)]$ and its logarithm, that is (\ref{r})-(\ref{echt}),
 thus gives
 the change of entropy as required.\\
A second answer is to look in the literature for motivated `general'
expressions for the entropy production, see \cite{chl}. There exists a
standard expression for the entropy production rate in stochastic Markov
dynamics.  For a finite state space and a nicely behaving continuous time
Markov chain with transition rate $k(\eta,\eta')$ for the probability per
unit time to change configuration $\eta$ to the new $\eta'$, we consider
\begin{equation}\label{take2}
\dot{S}(\rho) \equiv \sum_{\eta,\eta'} \rho(\eta) k(\eta,\eta') \ln
\frac{\rho(\eta)k(\eta,\eta')}{\rho(\eta')k(\eta',\eta)}
\end{equation}
for $\rho$ a measure on the state space.  This functional already
appeared in \cite{Sn}
 where a reference is made to Kirchhoff, \cite{Ki}, and it was
discussed again later, for example in the papers \cite{Q,QQ,els}.
  It is called the entropy production rate and we can verify
for our examples above or for other similar models that this is a good
name. Take say a hopping dynamics on the volume $V$ as in the asymmetric
exclusion process of example 3.1 but now with rates
\[
c(i,j,\eta) \equiv e^{-\frac 1{2}[H(\eta^{ij})-H(\eta) -
E_{ij}(\eta(i)-\eta(j))]}
\]
where $H(\eta)$ is the energy of $\eta$ and $E_{ij}=E_{ji}$ is an
external field.  Then, a small computation gives
\[
\dot{S}(\rho) = - \frac{d}{dt} \langle H(\eta_t)\rangle_\rho (t=0) +
{\cal P}_E  + \frac{d}{dt} \langle - \ln \rho(\eta_t)\rangle_\rho (t=0)
\]
which is the rate at which, in ``state'' $\rho$, energy is transferred to
the thermal reservoir (we have set the inverse temperature $\beta=1$) plus
the power ${\cal P}_E$ delivered by the external field  on the system plus
the rate at which $-\sum_\eta \rho(\eta) \ln \rho(\eta)$ (the Shannon
entropy) is changed. This is more or less okay and there are more nice
properties (like positivity, homogeneity and convexity of the functional
$\dot{S}(\rho)$, see \cite{els}).  There are however also bad things about this formula
for the entropy production rate.  First of all it is restricted to
Markovian stochastic dynamics. Secondly, it only gives the {\it mean}
entropy production in the state $\rho$ while it is not clear in what
sense it is the expectation of what variable quantity of which we could
study the (local) fluctuations.  Above all, it is bad because it seems
more or less useless if we do not know the state $\rho$, e.g. the
stationary state of the dynamics, to compute the steady state entropy
production (there is only an approximate minimum entropy production
principle to guide us here but it is of limited validity).\\
We can do better if we extend the formula to the space-time domain which
is exactly the set-up of (\ref{r})-({\ref{take1}); (\ref{take1}) is just
our alternative: we can show for the context of (\ref{take2}) that
\begin{equation}\label{takeup}
R^\tau(\rho) = \int_{-\tau}^\tau \dot{S}(\rho_t) dt
\end{equation}
the total entropy production over $[-\tau,\tau]$. Now that may seem more
complicated but there is in fact much gained. To see it, let us consider
the steady state.  In that case $\rho_t =\rho$ for all times and
$R^\tau(\rho) = 2\tau \dot{S}(\rho)$.  Now, $\rho$ is as unknown as
before but its extension to the path space measure $P=P_\rho^\tau$ is
much more accessible. $R^\tau(\rho)$ is the expectation of (\ref{echt})
and we can study its fluctuations and derive the symmetries discussed
before. Moreover, the formula for $R^\tau(\rho)$ is not at all restricted
to Markovian
dynamics; it is even not restricted to stochastic dynamics.\\
As a third general motivation, we give a Boltzmann-like space-time
counting interpretation of this entropy production. In order to give a
microscopic definition we present here only the simplest set-up that
mimics the start of equilibrium statistical mechanics but for space-time
trajectories.\\
Suppose we break up our stationary system into $N$ space-time cells which
are very small but still large enough to associate to each of them a
value for the current (the nature of which we do not need to specify
here).  We assume there are $n$ possible values $J_k$ with time-reversed
value $J_{k'} = -J_k$.
 The system is observed in these little windows and
we call the sequence of integers $(m_1,\ldots,m_n)$ with $m_1+\ldots +
m_n=N$ a distribution of current values; say, we find $m_k$ times out of
the $N$ observations the value $J_k$ so that $\sum_k m_k J_k$
approximates the space-time integrated current over our system. We will
also use the proportions $P_k \equiv m_k/N$. The distribution $(P_k)$
will be kept fixed and characterizing the nonequilibrium condition;
$\sum_k P_k J_k \equiv J \neq 0$ so that the distribution over the values
$(J_i))$ is not time-reversal invariant.  We forget about the origin
 of this distribution and it can be produced from surface or bulk driving conditions.\\
We select $M$ space-time cells from the interior of the space-time system
and we suppose that out of the corresponding $M$ observations we find
$r_k$ times the value $J_k$.  We also write $Q_k\equiv r_k/M$. Let
$W(r_1,\ldots,r_n;m_1,\ldots,m_n)$ denote the number of ways we can
achieve such a distribution.  It is given as a product of two multinomial
coefficients:
\[
W(r_1,\ldots,r_n;m_1,\ldots,m_n) = \frac{M!(N-M)!}{\prod_k r_k!
(m_k-r_k)! }
\]
Therefore, the relative weight for observing the sequence
$(r_1,\ldots,r_n)$ versus the sequence $(r_1',\ldots,r_n')$ for given
fixed distribution $(m_1,\ldots,m_n)$ is
\[
\frac{W(r_1,\ldots,r_n;m_1,\ldots,m_n)}
{W(r_1',\ldots,r_n';m_1,\ldots,m_n)} = \prod_k
\frac{r_k'!(m_k-r_k')!}{r_k!(m_k-r_k)!}
\]
If we take the limit of this expression for $N$ very large  while fixing
the distribution $(P_k)$ and we also let $M$ be very large, we obtain as
relative probabilities for two trajectories $\omega, \omega'$ of the
(internal) subsystem
\[
\frac{\mbox{Prob}[\omega]}{\mbox{Prob}[\omega']} = e^{M[S(Q)-S(Q') - \sum
(Q_k' -Q_k) \ln P_k]}
\]
with $S(Q)\equiv -\sum_k Q_k \ln Q_k$ and $\omega$ and $\omega'$
empirically corresponding to the distribution $(Q_k)$, respectively
$(Q_k')$.  We see therefore that those $\omega$ are most plausible that
have their $Q_k =P_k$. On the other hand, among two trajectories with
equal (equilibrium) entropy $S(Q)=S(Q')$, the one with larger dissipation
$\varphi(Q,P)\equiv \sum_k Q_k \ln P_k$ gets largest probability. Note
that we can rewrite the last term in the exponent above as
\[
\sum (Q_k' -Q_k) \frac{\partial S(P)}{\partial P_k}
\]
which is the product of the displacement $Q_k' - Q_k$ and the force
$\partial S(P)/\partial P_k$.  This term is not time-reversal invariant
if the distribution $(P_k)$ is not and in this way the condition of
microscopic reversibility is broken.  More precisely, by  taking $\omega'=
\Theta \omega$ where $\Theta$ again denotes time-reversal, we get
\[
\frac{\mbox{Prob}[\omega]} {\mbox{Prob}[\Theta \omega]} =  \exp M \sum_k
Q_k \ln \frac{P_k}{P_k'}
\]
where $P'$ is the time-reversal of the original $P$. In the exponent we
recognize the difference in dissipation
\[
\varphi(Q,P)-\varphi(Q',P)\equiv \sum_k Q_k \ln \frac{P_k}{P_k'}
\]
It is this function of the distribution $(Q_k)$ for the internal system
that we have called the (variable) entropy production
\[
R(Q,P) \equiv \sum_k Q_k \ln \frac{P_k}{P_k'}
\]
in the nonequilibrium state $P$.  Through the distribution $(Q_k)$, the
entropy production is also a function $R(\omega)$ of the space-time
trajectory $\omega$ and the fluctuations in the internal system will
always satisfy the symmetry
\[
\mbox{Prob}[R(\omega)=a] =\mbox{Prob}[R(\omega)=-a] \exp[Ma]
\]
on which we wrote before, see (\ref{flu}).\\
Of course, we will typically observe the mean entropy production
\[
\langle R \rangle \equiv \sum_k P_k \ln \frac{P_k}{P_k'}
\]
which is in fact the relative entropy  between the forward and the
backward nonequilibrium state that we introduced before.
   Furthermore, if we choose $P_k
\sim \exp[\lambda(J) J_k/2] $ with $\sum_k P_k J_k = J$ fixed, then this
becomes
\[
\langle R \rangle =  \lambda(J) J
\]
the standard product of force and current. The same scenario can be
applied to the case where coriolis or magnetic forces are present (but
then the time-reversal operation must take into account the parity of the
observables).

\section{Consequences}
Some of the general consequences of the Gibbsian hypothesis have already
been discussed in Section 2.2. and have been illustrated in Section 3.
Here we give some additional considerations.

\subsection{Local fluctuation symmetry}

The expressed symmetry in (\ref{flu}) first appeared in the context of
dynamical systems, in simulations with thermostated dynamics in
\cite{ecm}, for strongly
 chaotic systems in \cite{gc1,gc2,Gen,Ru4} and in stochastic dynamics in \cite{K,LeS,M,M4}.
An experimental verification was sought in \cite{cl}.
 The reason why it got so much attention is that there are indications
that it provides a far from equilibrium relation hopefully extending close
to equilibrium theory.  One can for example base on it a derivation of
Green-Kubo relations with the close to equilibrium Onsager
reciprocities, see \cite{G,G1,LeS,M}. Our way of looking at it is that it
plays a similar role to that of Ward identities in field theory. What is
new in (\ref{locr})-(\ref{lft}) is the local aspect of the symmetry
relation.  That makes more physical sense for spatially extended systems
because no fluctuations will ever be observed that take the size of the
whole macroscopic system.  This local aspect was already studied in
\cite{G3,GP,M} but now we have a systematic control over it. The results
that we obtained in \cite{mmj} for the asymmetric exclusion process
follow exactly the Gibbsian scheme.  The distribution $P$ is there the
restriction of
 the steady state process to the subvolume $\Lambda$.  This is a non-Markovian process
 but our Gibbsian set-up is verified and does not care about that. The corresponding
$R_\Lambda$ for this local process is, up to a space-time boundary term,
exactly equal to the local irreversible work and that is why we got a
local fluctuation theorem as explained in Section 3.1.\\
This set-up is very reminiscent of the Onsager-Machlup theory except that
there one starts from a linear dynamics for the macroscopic variables.
The bulk of their paper is the construction of the associated action
functional $A$.  All that is said here is that we can do the same
starting from a non-linear dynamics on a more microscopic level for the
dynamics.  Then $A$ is not quadratic but the fluctuation symmetry still
holds.

\subsection{Positivity of entropy production}

From (\ref{pos}) it is rather easy to obtain the positivity of the entropy
production.  When this is a simple product of field and current, it will
give also the average direction of the current, as in (\ref{poso}).  For
example, in the example of heat conduction above, it is easy to establish
that the heat current will flow from the hot towards the cold reservoir.
That is not new, see \cite{eck1,eck2,l1}, but our approach seems more
general.  In particular,
 all this remains largely unchanged
for non-Markovian dynamics, \cite{mm}.\\
What is not so clear {\it a priori} is whether this situation is stable
under taking thermodynamic limits. That is, whether we could have a
non-vanishing current even when there is no entropy production. The
answers are given in \cite{M2,M3,M5}.
 The opposite is rather easy; you
 can have a zero net current and still $\langle R \rangle > 0$, see an example in
 \cite{M4}.

\subsection{Response relations}
We restrict us here to nonequilibrium steady states. One of the most
important consequences of (\ref{af}) is that we can express the
expectation of a function in the stationary state $\rho$ (at a fixed
time) as the expectation in a Gibbsian ensemble
\[
\int f(\eta) d\rho(\eta) = \int dP(\omega)  f(\omega_0)
\]
because $\rho$ is the restriction of the steady state distribution $P$ to
a(ny) fixed time layer. As already announced after (\ref{flu}) we can in
principle derive exact identities between derivatives of space-time
correlation functions via the differentiation of the Ward identity.  That
was already explored in \cite{G,LeS,M} but it has not been made out
whether for example a standard Green-Kubo relation as it should appear in
Fourier's law, is within reach now, see \cite{Fou}.  We will not repeat
here the scheme of the approach starting from (\ref{flu}) or from the
fluctuation symmetry but instead, we give here an idea of how close we
can get via a direct Gibbsian approach. One should compare it with the
usual formal perturbation theory applied to the nonequilibrium dynamics,
\cite{Fou}.\\
We have a steady state $P$ and a stationary measure $\rho$ which is
obtained by perturbing an equilibrium dynamics with corresponding $P_o$
and $\rho^o$.  Let us denote by $\varepsilon$ the small parameter.  For
example 3.2, in the case of heat conduction in a crystal with stochastic
reservoirs, $\varepsilon \equiv (T_\ell -T_r)/2$ with
$T_r=1/\beta_r,T_\ell=1/\beta_\ell$.  We also write $T\equiv (T_\ell
+T_r)/2=1/\beta$. Take a function $f$ that is antisymmetric under
time-reversal, $f\Theta = -f$.  A good example is a current, e.g. the sum
of heat currents $J_k$ of (\ref{cur}) at the oriented bonds $\langle
k-1,k\rangle\subset L_N$ in our crystal: we take some $0<M=M(N)<N$ and put
\[
f(\omega) = \frac 1{N} \sum_{-M}^{M} J_k(0)
\]
 The
idea is now very simple.  Simply write the expectation in the stationary
state as
\[
\int f(p,q) \rho(p,q) dpdq = \int dP(\omega) f(\omega)
\]
where $P$ is as always the steady state process over a time interval
$[-\tau,\tau]$. This is indeed very cheap; all what happens is that we
imbed the stationary distribution in the larger Gibbsian distribution $P$.
$P$ depends on $\varepsilon$ and for $\varepsilon=0$ it just equals
$P_o$, our reference process. Yet, now we can apply (\ref{af}) and write
\[
\int f(p,q) \rho(p,q) dpdq = \int dP_o(\omega) e^{-A(\omega)} f(\omega)
\]
and we can take the derivative with respect to $\varepsilon$ at
$\varepsilon=0$ to find the linear response behavior.  To be specific we
turn to the formula (\ref{girs}), (\ref{H}) and (\ref{rheat}).  The
computation gives
\begin{eqnarray}\label{lo}
&&\frac{d}{d\varepsilon}\int f(p,q) \rho(p,q) dpdq \,(\varepsilon=0)=
\frac{d}{d\varepsilon} \int \frac{dP_o(\omega)}{\rho^o(\omega_{-\tau})}
\rho(\omega_{-\tau}) f(\omega)\,(\varepsilon=0)\nonumber \\&+&\frac
1{2T^2} \int dP_o(\omega) f(\omega) \{[\frac 1{2}p_{-N}^2(\tau) - \frac
1{2} p_{-N}^2(-\tau)+ \int_{-\tau}^\tau \frac{\partial U_N}{\partial
q_{-N}}(q(t)) p_{-N}(t)dt] \nonumber \\&-& [\frac 1{2} p_{N}^2(\tau) -
\frac 1{2} p_N^2(-\tau) +  \int_{-\tau}^\tau \frac{\partial U_N}{\partial
q_{N}}(q(t)) p_{N}(t)dt]\}
\end{eqnarray}
We can still rewrite this using (\ref{sst}) as
\begin{equation}\label{i1i2}
\frac{d}{d\varepsilon}\int f(p,q) \rho(p,q) dpdq \,(\varepsilon=0)= I_1 +
I_2
\end{equation}
with
\[
I_1 \equiv \frac{d}{d\varepsilon} \int
\frac{dP_o(\omega)}{\rho^o(\omega_{-\tau})} \rho(\omega_{-\tau})
f(\omega)\,(\varepsilon=0) - \frac{d}{d\varepsilon} \int
\frac{dP_o(\omega)}{\rho^o(\omega_{-\tau})}\bar\rho(\omega_{-\tau})]f(\omega)\,(\varepsilon=0)
\]
where
\[
\bar\rho(p,q)=\frac 1{Z} \exp -[\beta_\ell H_N^{< 0}(p,q) +\beta_r
H_N^{\geq 0}(p,q)]
\]
is a two-temperature Gibbs measure for $H_N^{<0}(p,q)\equiv
\sum_{-N}^{-1}[p_i^2/2 + V_i(q_i) + \lambda_i\Phi(q_i-q_{i+1})] \equiv
H_N - H_N^{\geq 0}$, and
\[
I_2\equiv \frac 1{NT^2} \sum_{k=-M}^M\int dP_o(\omega) \int_{-\tau}^\tau
dt \, J_k(0)J_0(t)
\]
Equation (\ref{i1i2}) is valid for all $\tau,N$ and $M$ and it describes
the linear response.  At this moment, it would be interesting to consider
the various possible limits (and possible exchanges with the
$\varepsilon$-derivative) but we can only make some general comments.
The first term $I_1$ is the $\varepsilon-$derivative of a difference of
two expectations. The expectations are under the equilibrium dynamics of
the current $f$ at time $0$ started at time $-\tau$ in the nonequilibrium
stationary state $\rho$, respectively the nonequilibrium state
$\bar\rho$. Even though there is a non-vanishing current at time $-\tau$,
we expect that for sufficiently large $\tau$ the current will be
arbitrarily small at time $0$: the expectation of an antisymmetric
function will go to zero under the equilibrium dynamics.  The remaining
term $I_2$, of the form $I_2 = \kappa/N$, should give rise to a Green-Kubo formula for
the thermal conductivity $\kappa$, see e.g. \cite{sp}, in the
limits $N, \tau\uparrow +\infty$. Note however that the reservoirs are still
present in the time-evolution for $J_0(t)$. It appears therefore that the
correct limit is to first take $M,N\uparrow +\infty$ before
$\tau\uparrow+\infty$ in the equilibrium current-current correlation
function. In this way the bulk Hamiltonian dynamics, having a finite
horizon of propagation, should dominate the time-evolution and the
correct Green-Kubo formula should appear, see \cite{Fou}. This however
requires an analysis that goes much beyond the generalities that have
been the subject of the present paper.  After all, the above rigorous derivation
for finite space-time extension holds also in the harmonic case, while we
know that there the current $\langle J_0 \rangle$ is proportional to the temperature difference
$\varepsilon$
(non-vanishing as the size $N$ of the system tends to infinity).  The
problem of estimating the current-current correlations in the thermodynamic limit remains therefore
essential and open.

\bigskip
\noindent{\bf Acknowledgment:} I have taken advantage of an ongoing
collaboration with Karel Neto\u cn\' y, Frank Redig and Michel
Verschuere.  I am also grateful to J. Bricmont, E. Cohen, G. Gallavotti,
S. Goldstein, J. Lebowitz and H. Wagner for discussions.

\end{document}